\documentclass[showpacs,preprintnumbers,amsmath,amssymb,12pt]{revtex4}

\usepackage{dcolumn,amsmath,xspace}

\usepackage{graphicx}
\usepackage{dcolumn}
\usepackage{amsmath}

\usepackage{units}

\newcommand{\degC}{\ensuremath{\!^{\circ}\text{C}}}
\newcommand{\sixrt}{\ensuremath{(6\sqrt{3}\!\times\!6\sqrt{3})\text{R}30}~}

\begin{document}
\title{The interface structure of epitaxial graphene grown on 4H-SiC(0001)\\}

\author{J. Hass}
\author{J.E. Mill\'{a}n-Otoya}
\author{P. N. First}
\author{E. H. Conrad}
\affiliation{The Georgia Institute of Technology, Atlanta, Georgia 30332-0430, USA}

\begin{abstract}
We present a structural analysis of the graphene-4HSiC(0001) interface using surface x-ray reflectivity. We find that the interface is composed of an extended reconstruction of two SiC bilayers.  The interface directly below the first graphene sheet is an extended layer that is more than twice the thickness of a bulk SiC bilayer ($\sim\!1.7$\AA~ compared to 0.63\AA). The distance from this interface layer to the first graphene sheet is much smaller than the graphite interlayer spacing but larger than the same distance measured for graphene grown on the $(000\bar{1})$ surface, as predicted previously by {\it ab intio} calculations.
\end{abstract}
\vspace*{4ex}

\pacs{68.55.-a, 68.35.-p, 61.10.Kw, 61.46.-w}
\keywords{Graphene, Graphite, X-ray reflectivity, X-ray
  diffraction, SiC, Silicon carbide}
 \maketitle
\newpage

\section{Introduction}
The direct growth of single or multiple sheets of graphene on an insulating or semiconducting substrate is known as epitaxial graphene (EG). Because this material has been identified as a viable all-carbon candidate for post CMOS electronics,\cite{ITRS_2007} there is a strong impetus for the study of EG and how it can be produced, lithographically patterned and made into electronic devices.\cite{Berger04,Berger06,de Heer_SSC_07,Hass_APL_06,Kedzierski_transistors} The current substrate for EG growth is either of the two polar faces of hexagonal SiC: the SiC(0001) Si-terminated surface (Si-face) and the SiC$(000\bar{1})$ C-terminated surface (C-face). Multilayer epitaxial graphene films grown on C-face SiC show electronic properties expected for an isolated graphene sheet including a Berry's phase of $\pi$, weak anti-localization, and a square root dependence of the Landau level energies with applied magnetic field.\cite{Berger06,Sadowski_PRL_06,Sadowski07a,Wu07,de Heer_SSC_07} The fact that transport in graphene grown on SiC substrates is so similar to transport properties expected for an isolated graphene sheet is quite remarkable since graphene/substrate interactions and multilayer graphene stacking should potentially influence the 2D Dirac electrons responsible for the unusual properties of graphene. On the other hand, these findings are very fortuitous and have made graphene grown on SiC the focus of research targeting a path towards graphene electronics.

Studies of the graphene/SiC substrate and multilayer graphene stacking have begun to elucidate the relationship between EG's electronic properties and its structure. It is very clear from the earliest measurements that both the growth and structure of C-face and Si-face grown graphene are very different.\cite{vanBommel75,Muehlhoff_JAP_86,Forbeaux_PRB_98,Forbeaux_SS_99,Forbeaux00,Hass_APL_06} Graphene prepared on SiC in ultra high vacuum (UHV) grows relatively slowly on the Si-face as compared with the C-face~\cite{Muehlhoff_JAP_86} and tends to be $\sim\!1\!-\!5$ layers thick, while graphene grown on the C-face grows rapidly at lower temperatures compared to the Si-face and can achieve thickness well above 5 layers.\cite{Forbeaux_PRB_98,Forbeaux_SS_99,Hass_APL_06} In addition, graphene grown on the C-face in UHV contains a high concentrations of graphene nanocaps~\cite{Naitoh_SRL_03} and other defects,\cite{Varchon_PRB_08} while graphene growth on the Si-face causes the SiC substrate to roughen.\cite{Hass_APL_06} Furnace grown C-face graphene on the other hand reaches domain sizes $>\!5\mu$m,\cite{Hass_JPhyCM_08} many orders of magnitude larger than those grown in UHV on either the C-face or Si-face.\cite{Hass_JPhyCM_08} Correspondingly, electron mobilities and electronic coherence lengths are an order of magnitude higher on the furnace grown C-face graphene compared to UHV grown Si-face graphene.\cite{Berger06,Hass_APL_06} Finally, Si-face graphene is epitaxial with \sixrt periodicity as observed by low energy electron diffraction (LEED) (i.e, the graphene is rotated $30^\circ$ relative to the SiC $<\!10\bar{1}0\!>$ direction).\cite{vanBommel75,Forbeaux_PRB_98} C-face films, on the other hand, can have multiple orientational phases~\cite{vanBommel75,Forbeaux00,Hass_PRL_08,Varchon_PRB_08} with a unique stacking order.\cite{Hass_PRL_08,Hass_JPhyCM_08}

Because of the relative ease of growing graphene in UHV on SiC(0001), the Si-face has been the most extensively studied of the two hexagonal polar surfaces.\cite{Owman_SS_95,Chang_SS_91,Li_SS_96,Charrier_JAP_02,Johansson_PRB_96,Brar_APL_07,Rollings_06,Ohta_Sci_06,Seyller_SS_06,Rutter_Sci_07,Rutter_PRB_07,Riedl_PRB_07,Hibino_PRB_08,Ohta_LEEM} Despite this body of work, the structure of this interface is still unknown even though it is crucial for understanding the electronic structure of subsequent graphene layers and the charge transfer between the substrate and the graphene film.
Early studies of the graphitization of the Si-face presumed that the \sixrt pattern was due to the commensurate alignment of a graphene overlayer with an unreconstructed SiC surface, forming a moir\'{e} pattern.\cite{vanBommel75,Chang_SS_91,Li_SS_96,Charrier_JAP_02} However, the consensus from recent experiments by many groups is that the \sixrt structure is a true structural precursor phase to graphene formation that persists, although slightly altered, after graphene has formed.\cite{Johansson_PRB_96,Rutter_PRB_07,Riedl_PRB_07,Brar_APL_07,Lauffer_PRB_08} While scanning tunneling microscopy (STM),\cite{Owman_SS_95,Ong_PRB_06,Rutter_PRB_07,Riedl_PRB_07,Brar_APL_07,Mallet_PRB_07} has outlined some of the lateral features of the \sixrt structure, the buried interface structure remains elusive.

In this paper, we use specular x-ray reflectivity to measure the buried structure of the graphene/4H-SiC(0001) interface. We find that the first layer of carbon with an areal density of graphene sits close to the last layer of atoms in the interface layer. For the Si-face the graphene-interface spacing is found to be $2.32\pm 0.08\text{\AA}$. This number is consistent with {\it ab-initio} calculations that predict a covalently bonded insulating first graphene layer that ``buffers'' subsequent graphene layers from the substrate.\cite{Varchon_PRL_07,Mattausch_PRL_07} However, unlike the simple bulk terminated interface used in these calculations, the interface layer is found to be more complex. The structure of this interface layer suggests it plays an important role mediating the interaction of the SiC substrate with the graphene film.

\section{Experimental}
The substrates used in these studies were 4H-SiC purchased from Cree, Inc.\cite{Cree} Prior to graphitization the $3\!\times\!4\!\times\!0.35$mm samples were ultrasonically cleaned in acetone and ethanol. The Si-face samples were $\text{H}_2$ etched\cite{Hass_JPhyCM_08} and subsequently graphitized in UHV ($P<\unit[1\times10^{-10}]{Torr}$) by electron-bombardment heating. Substrates were first heated to \unit[1100]{\degC} for \unit[6]{min} to form a $(\sqrt{3}\times\!\sqrt{3})$R30 reconstruction, after which they were heated to \unit[1320]{\degC} for \unit[8]{min} to remove surface contamination and to form a well ordered $(6\sqrt{3}\times\!6\sqrt{3})$R30 reconstruction.  The samples were subsequently heated to \unit[1400--1440]{\degC} for \unit[6--12]{min} to create graphene films 1--3 layers thick. During graphitization the pressure in the system reached $P\lesssim\unit[1\times10^{-8}]{Torr}$.  Despite the high pressure during growth, sample order is indistinguishable from Si-face sample order reported by other groups who grow at pressures $\sim\!10^{-10}$torr or from those who use a Si flux to reduce surface oxides.\cite{Owman_SS_95,Rutter_Sci_07,Brar_APL_07,Charrier_JAP_02,Hass_APL_06} Growth-induced substrate roughening in all reported investigations of Si-face graphitization leads to average SiC terrace sizes of $\sim\!\unit[300-500]{\AA}$ (graphene order is actually bigger than the terrace size because graphene grows over the SiC steps.\cite{Seyller_SS_06}) It appears that the long range order of UHV grown graphene on the Si-face is not very sensitive to most details of the surface preparation.

Once the samples were graphitized, they remained inert allowing them to be transported into the separate x-ray scattering chamber for analysis. The x-ray scattering experiments were performed at the Advanced Photon Source, Argonne National Laboratory, on the 6IDC-$\mu$CAT UHV ($P < \unit[2\times10^{-10}]{\text{torr}}$) beam line. Experiments were performed at a photon energy of $16.2~$keV. The number of graphene layers on the sample was determined by both Auger electron spectroscopy (AES) and x-ray reflectivity (as described in the next section). The reflectivity data is presented in standard SiC hexagonal reciprocal lattice units [{\it r.l.u.}] $(h,k,\ell)$. These are defined by the momentum transfer vector ${\bf q}\!=\!(h{\bf a}^*_\text{SiC},k{\bf b}^*_\text{SiC},\ell{\bf c}^*_\text{SiC})$, where $a^*_\text{SiC}\!=\!b^*_\text{SiC}\!=\!2\pi/(a_\text{SiC}\sqrt{3}/2)$ and $c^*_\text{SiC}\!=\!2\pi/c_\text{SiC}$) [$a_\text{SiC}$, $b_\text{SiC}$ and $c_\text{SiC}$ are the standard 4H-SiC hexagonal lattice constants]. The measured lattice constants were $a_\text{SiC}\!=\!3.079\pm 0.001\text{\AA}$, $c_\text{SiC}\!=\!10.081\pm .002\text{\AA}$ and are within error bars of published values.\cite{Bauer_PRB_98} For reference the nominal hexagonal lattice constants for graphite are $a_\text{G}\!=\!2.4589\text{\AA}$, $c_\text{G}\!=\!6.708\text{\AA}$.\cite{Baskin_PR_55}

\section{Results}
To obtain detailed information about both the graphene films and the SiC-graphene interface, we have measured the surface x-ray specular reflectivity from graphitized 4H-SiC(0001). Details of the data collection and the model are similar to those used to determine the structure of the 4H-SiC$(000\bar{1})$/graphene system.\cite{Hass_PRB_07} The data is collected by integrating rocking curves for different perpendicular momentum transfer vectors, $q_\perp=2\pi\ell/c_\text{SiC}$, where ${\bf q}\!=\!{\bf k}_f\!-\!{\bf k}_i$. Since the reflectivity only depends on $q_\perp$ (or equivalently $\ell$), the data can be analyzed using a one-dimensional model where all lateral information is averaged over the $0.4\!\times\!0.4$mm x-ray beam.

\begin{figure}
\begin{center}
\includegraphics[width=10.0cm,clip]{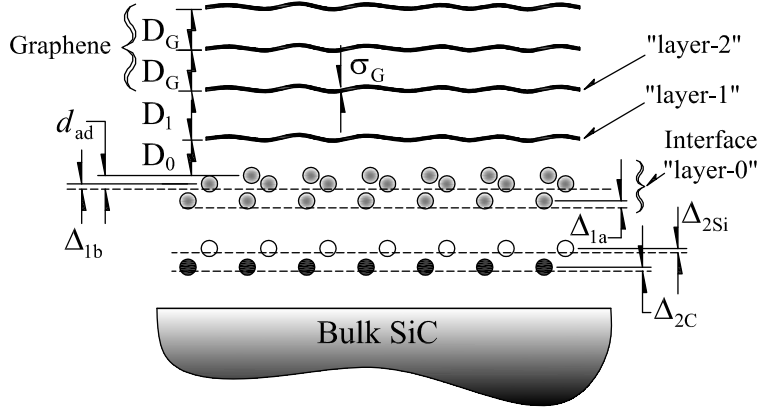}
\end{center}
\caption{A schematic model of multi-layer graphene grown on the 4H-SiC(0001) substrate. Dashed lines are the bulk SiC lattice planes before interface relaxation ($\Delta$'s). The $5^\text{th}$ plane of atoms (adatom) is displaced $d_{ad}$ from the topmost atom plane in the interface. ($\bullet$) are carbon atoms and ($\circ$) are silicon atoms. The shaded circles in the interface (``layer-0'') can be either carbon or silicon atoms. The graphene layers above the interface layer are referred to as ``layer-'', -2, -3, etc.}
\label{F:Si-model_sch}
\end{figure}

The schematic model of the graphene covered SiC(0001) surface is shown in Fig.~\ref{F:Si-model_sch}. This general model allows us to explore a number of possible graphene/SiC structures proposed by previous experiments. In the model the SiC substrate contribution is broken into two terms: (i) the amplitude from a bulk terminated surface, (ii) the combined amplitude from a relaxed SiC bilayer just above the bulk and a reconstructed interface layer (defined as the interface ``layer-''). The scattered x-ray intensity $I(\Theta,\ell)$ is then the result of a sum of three scattered amplitudes; the bulk $F_\text{bulk}$, the interface region $F_{I}$, and the graphene $F_\text{G}$:

\begin{eqnarray}
I(\Theta,\ell)&=& A(\Theta,\ell) e^{-4\gamma_{\text{SiC}}\sin^2{\pi \ell/2}}\nonumber\\
 \times &&\left|\frac{F_\text{bulk}(\ell)}{1-e^{-2\pi{i\ell}}} + F_{I}(\ell) + \frac{\rho_{\text{G}}}{\rho_{\text{SiC}}}F_\text{G}(\ell)\right|^2.
\label{E:intenisty}
\end{eqnarray}
$A(\Theta,\ell)$ is a term that contains all corrections due to the experimental geometry.\cite{Robinson_review,Vlieg_JAC_97,Feng_thesis} The exponential term accounts for the substrate roughness caused by half-cell step fluctuations in the SiC surface (the predominant step height on 4H samples;\cite{Hass_APL_06} $c_{SiC}/2$). $\gamma_{\text{SiC}}$ is the variance in the number of half-cell layers in the surface due to steps.\cite{Elliott_PYSICAB_96} Roughly, $\gamma_{\text{SiC}}$ is proportional to the SiC step density. The first term in Eq.~(\ref{E:intenisty}) is the bulk 4H-SiC structure factor, $F_\text{bulk}(\ell)$,\cite{Bauer_ACryst_01} modified by the crystal truncation term, $(1-e^{-2\pi{i\ell}})^{-1}$ (Ref.~\onlinecite{Robinson_CTR_86}). $F_\text{G}(\ell)$ in Eq.~(\ref{E:intenisty}) is weighted by the ratio of the areal densities of a 4H-SiC$(0001)$ and a graphene (0001) plane; $\rho_{\text{G}}/\rho_{\text{Si}}$=3.132, to properly normalize the scattered amplitude from the graphene layer per 4H-SiC$(0001)$ $(1\!\times\!1)$ unit cell.

$F_{I}(\ell)$ in Eq.~(\ref{E:intenisty}) is the structure factor of the interface region between the bulk and the graphene film. Although we cannot obtain lateral information about the SiC(0001) \sixrt structure from reflectivity data, the vertical shifts of atoms and layer density changes associated with them can be determined. To begin to understand this interface, we allow for a reconstruction by placing a SiC bilayer plus an interface containing up to three additional atomic layers between the bulk and the multi-layer graphene film [see Fig.~\ref{F:Si-model_sch}]. We then write the interface structure factor as:
\begin{equation}
F_{I}(\ell)=\sum_{j=1}^5{f_j(\ell )\rho_je^{i2\pi\ell z_j/c_\text{SiC}}},
\label{E:Surf_form}
\end{equation}
where $\rho_j$ is the relative atom density for the $j^{\text{th}}$ interface layer ($\rho_j=1$ for a bulk layer corresponding to $8.22\!\times\!10^{-16}\text{atoms cm}^{-2}$) at a vertical position $z_j$ (the zero height is chosen as the last layer of atoms in the interface). $f_j(\ell )$ is the atomic form factor of C or Si. The fifth atom layer is added to explore the possibility of adatoms between the SiC and the graphene.

To be completely general the scattered amplitude from the graphene film takes into account the possibility of a lateral distribution of varying graphene layers. This is done by defining an occupancy parameter $p_n$ as the fractional surface area covered by all graphene islands that are $n$ graphene layers thick. $p_n$ is subject to the constraint equation $\sum{p_n}=1$, where $p_0$ is the fraction of area that has no graphene. The multilayer graphene structure factor can then be written in the general form:

\begin{subequations}
\label{E:island_I}
\begin{multline} F_{G}(\ell) = f_{C}(\ell )\sum_{n=1}^{N_{\text{max}}}p_{n}\left\{
\sum_{m=1}^{n}F_m(\ell)e^{2\pi{il}z_m/c_\text{G}}\right\},
\label{E:Fpb}
\end{multline}
\begin{equation}
z_m=\left\{\begin{array}{ll}
D_0 + (m-1)D_1 &\mbox{$m \leq 2$} \\
D_0 + D_1+(m-2)D_\text{G} &\mbox{$m > 2$}\label{E:beg4}
\end{array}
\right. .
\end{equation}
\label{E:grap_stru}
\end{subequations}
$f_{C}$ is the atomic form factor for carbon and $N_{\text{max}}$ is the number of layers in the thickest graphene film on the surface. $D_0$ is the spacing between the bottom layer of an island and the last atom layer in the interface. $D_1$ is the spacing between graphene layer-1 and layer-2, while $D_\text{G}$ is the average layer spacing between graphene in subsequent layers [see Fig.~\ref{F:Si-model_sch}].

Because STM studies of multi-layer graphene films grown on the Si-face indicate some buckling of the graphite layer,\cite{Owman_SS_95,Riedl_PRB_07,Rutter_PRB_07} we must allow for a small vertical height distribution in each graphene layer that gives rise to a structure factor, $F_m(\ell)$ in Eq.~(\ref{E:Fpb}), for each layer. A vertical modulation of the graphene layers can be modeled two ways.  The simplest method is to assume an average, layer independent, random vertical disorder, $\sigma_\text{G}$, that will give rise to a Debye-Waller term for each layer, i.e. $F_m(\ell)=  e^{-q_\perp^2\sigma_\text{G}^{2}/2}$. Because the vertical modulation is known to decay quickly after the first graphene layer,\cite{Riedl_PRB_07} a more refined model uses the same average Debye-Waller term for the upper graphene layers but allows for a different distribution of carbon atoms in the first graphene layer such as those calculated in {\it ab intio} calculations.\cite{Varchon_axCM_Ripples} In this case the the structure factor of the first layer, $F_1(\ell)$, needs to be known.  As we'll show, both models give very similar results.

\begin{figure}
\begin{center}
\includegraphics[width=10.0cm,clip]{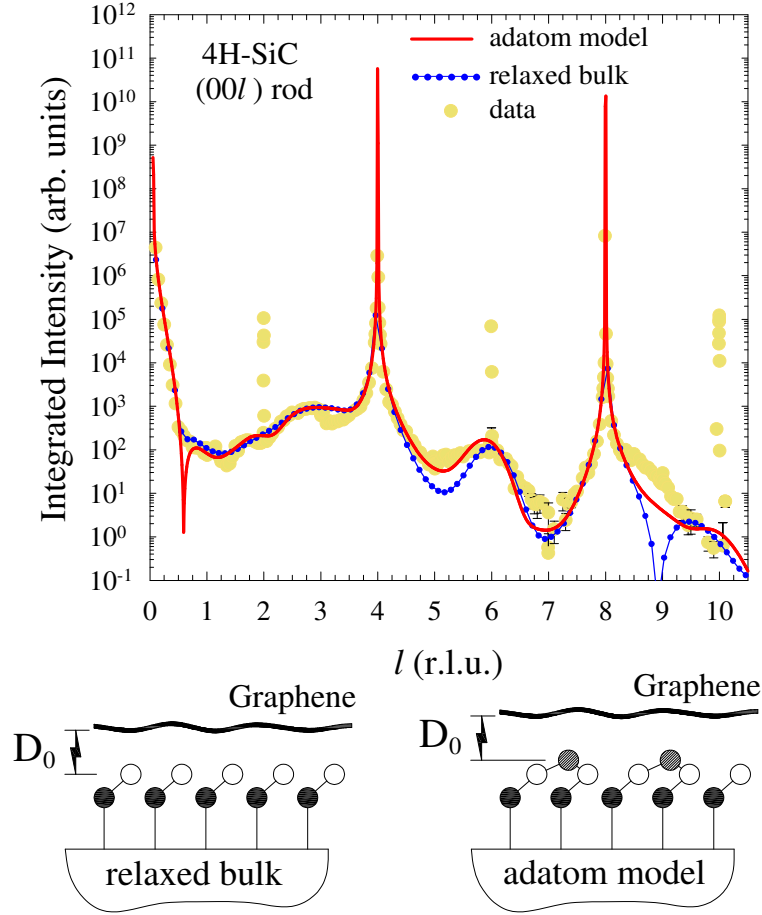}
\end{center}
\caption{(color online) Specular reflectivity versus $q_\perp$ (in r.l.u.) for a graphitized 4H-SiC(0001) Si-face surface.  Circles are the data. Fits to the two model structures in the figure are given. In the schematic models filled and open circles are C and Si atoms, respectively. Shaded circles are Si adatoms. Dotted blue line is a fit to a bulk terminated SiC(0001) surface with a single relaxed bilayer. Solid red line is a fit to a model similar to the relaxed bilayer but with the addition of a layer of Si-adatoms with $\rho_\text{ad}\!=0.21$.} \label{F:reflec_1}
\end{figure}

Reflectivity data for a Si-face multi-layer graphene film is shown in Fig.~\ref{F:reflec_1}. The main bulk 4H-SiC peaks occur at $\ell\!=\!4$ and $\ell\!=\!8$.  The sharp peaks at $\ell\!=\!2$, 6 and 10 are the ``quasi-forbidden'' reflections of bulk SiC.\cite{Bauer_ACryst_01} In SiC reciprocal lattice units, the graphite bulk reflections are nominally expected at $\ell\!\sim\!3$, 6 and 9 (i.e., $\ell\!=\!\ell_\text{G}(c_\text{SiC}/c_\text{G})$, where $\ell_\text{G}$ = 0, 2, 4 etc.).

We have tested a number of structural models for the graphene/4H-SiC$(0001)$ interface.  While the majority of experimental studies point to a complicated interface structure, simple models consisting of a nearly bulk terminated substrate with the graphene on top are still being proposed.\cite{Emtsev_PRB_08} However, such models always give poor fits to the x-ray data. This is demonstrated in Fig.~\ref{F:reflec_1} where we plot the best fit reflectivity for a bulk terminated surface where the $6\sqrt{3}$ interface layer is essentially a graphene layer as suggested by Emtsev et al..\cite{Emtsev_PRB_08} In this model, the interface layer-0 in Fig.~\ref{F:Si-model_sch} is replaced by a single carbon layer with a graphitic density. The last bulk bilayer density is kept constant at the bulk value while the bilayer spacings are allowed to relax (the relaxation from the bulk value are small; $\Delta_\text{2C}\!=\!-0.03$\AA and $\Delta_\text{2Si}\!=\!0.01$\AA). All other parameters in Eqs.~(\ref{E:Surf_form}) and (\ref{E:grap_stru}) are allowed to vary to achieve the best fit shown in Fig.~\ref{F:reflec_1}. This includes the distance between the last SiC bilayer and the graphitic interface layer, $D_0$, and the distance between this layer and the next graphene layer, $D_1$, that relax to best fit values of 2.55\AA and 3.62\AA), respectively. Note that $D_1$ is larger than the bulk graphite spacing of 3.354\AA. As can be seen in Fig.~\ref{F:reflec_1}, this model give a very poor fit to the data at values of $\ell =$ 5 and 9, the anti-Bragg points for SiC. This is typical of all bulk terminated models including those with a substantial modulation of the first graphene layer such as the calculated \sixrt surface of Varchon et al.\cite{Varchon_axCM_Ripples} and Kim et al.\cite{Kim_PRL_08} The calculated reflectivity from this theoretical interface, including a relaxed bulk surface and the structure factor $F_1(\ell)$ of the rippled first graphene layer, gives similarly poor fits to the reflectivity near the anti-Bragg positions.

Better fits can be obtained by an extended interface where an additional partial layer of adatoms is added to the simple relaxed bilayer model [see the schematic models in  Fig.~\ref{F:reflec_1}].  As demonstrated in Fig.~\ref{F:reflec_1}, the additional density from the adatoms begins to correct many of the deficiencies in the relaxed bulk model fit at the anti-Bragg points (especially near $\ell\!=\! 9$). Note that the adatom model used to fit the reflectivity is very similar to the model proposed by Rutter et al.\cite{Rutter_PRB_07}, including the Si adatom density which is $\rho_\text{ad}\!=\!0.21$ compared to 0.22 in their model.

The improvement in the calculated intensities by adding an adatom layer is due to the increased scattered intensity at the SiC anti-Bragg condition that would be zero for a bulk terminated interface. Regardless, the simple adatom model cannot reproduce a number of features in the reflectivity data for $\ell\!>\!4$. The inability of this model to fit the experimental data is a result of both an insufficient atomic density in the interface and the atomic gradient through the interface. Increasing the width of the interface adds an additional Fourier component in Eq.~(\ref{E:Surf_form}) that both broadens the fit near $\ell\!\sim\!6$ and removes the interference minimum at $\ell\!\sim\!9$. Therefore, to improve the fits, it is necessary to change both the atom distribution and the thickness of the interface layer. The need for an additional plane of atoms is also consistent with number of STM experiments of the SiC(0001) \sixrt interface. STM images of ``trimer-like'' structures suggest at least one additional partial layer of atoms.\cite{Owman_SS_95,Ong_PRB_06,Rutter_PRB_07,Riedl_PRB_07,Brar_APL_07}

\begin{figure}
\begin{center}
\includegraphics[width=10.0cm,clip]{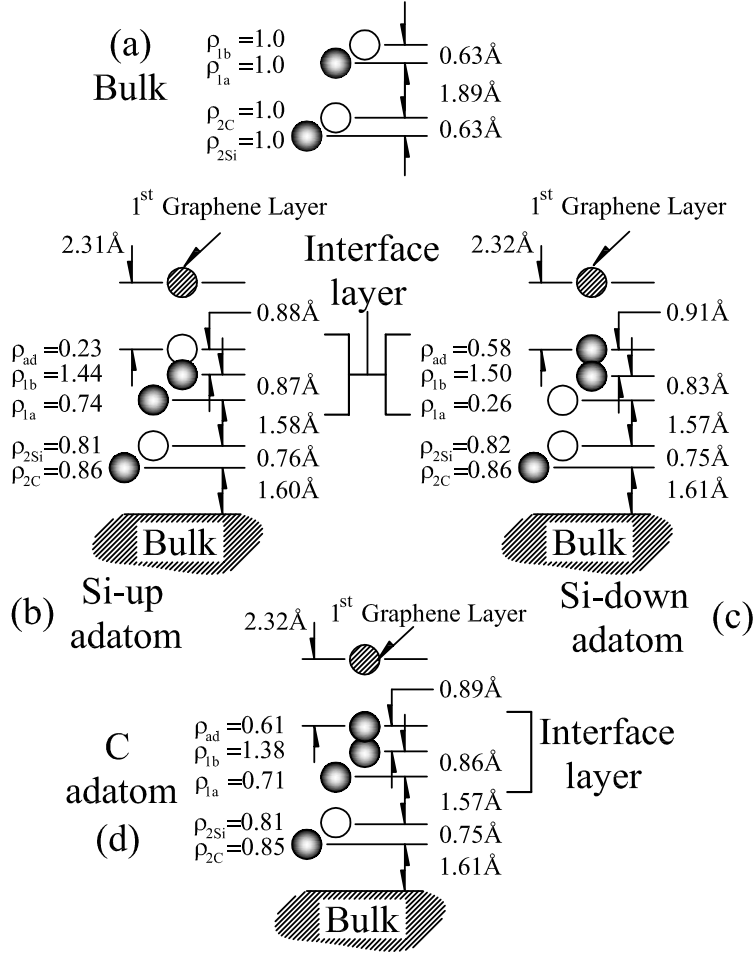}
\caption{Three graphene/SiC interface ball models for graphene grown on the Si-face of SiC determined by surface x-ray reflectivity; (b) Si-up model, (c) Si-down model and (d) C-adatom model. Open circles are silicon atoms and shaded circles are carbon atoms. The densities $\rho$ are relative to the densities of bulk SiC shown in (a).} \label{F:Si-facemodel}
\end{center}
\end{figure}

Adding a $5^\text{th}$ layer of atoms and changing the atom density in the interface layer-0 leads to a set of nearly identical structures that are shown in Fig.~\ref{F:Si-facemodel}. These structures are distinguished by whether an atom plane in the interface is composed of either carbon or silicon atoms. This is because in x-ray diffraction the ratio of atomic form factors of Si and C used in Eq.~(\ref{E:Surf_form}) is determined, to first order, by the ratio of their atomic numbers $14/6=2.33$. Therefore, the model calculation should give a similar fit if all silicon atoms are replaced by carbon atoms with 2.33 times the density (however, there is a substantial difference in the $\ell $ dependence of the Si and C atomic form factors that affects both the $\rho$'s and $z_j$'s in the final fits). While x-ray reflectivity data alone is unable to distinguish between different silicon and carbon compositions in the interface layer, spectroscopic data from a number of experimental groups place restrictions on the atomic makeup of layer-0 in Fig.~\ref{F:Si-model_sch}.

Angle resolved photoemission spectroscopy (ARPES) studies by Emtsev et al.\cite{Emtsev_MSF_07} as well as x-ray photoemission spectroscopy (XPS) studies by Johansson et al.\cite{Johansson_PRB_96} conclude that the interface layer has a significant carbon concentration (at least 1.3 times more than the carbon in a bulk SiC bilayer~\cite{Johansson_PRB_96}) that rules out a purely silicon interface. In fact the x-ray diffraction also rules out a purely silicon interface because the density of silicon required to get reasonable fits to the reflectivity data is almost half the density required for an $\text{sp}^3$ silicon film, which is physically unreasonable. These spectroscopic constraints reduce the number of possible structures that are compatible with the reflectivity data to three: the ``C-adatom'', ``Si-up'', and ``Si-down'' models shown in Fig.~\ref{F:Si-facemodel}.

\begin{figure}
\begin{center}
\includegraphics[width=10.0cm,clip]{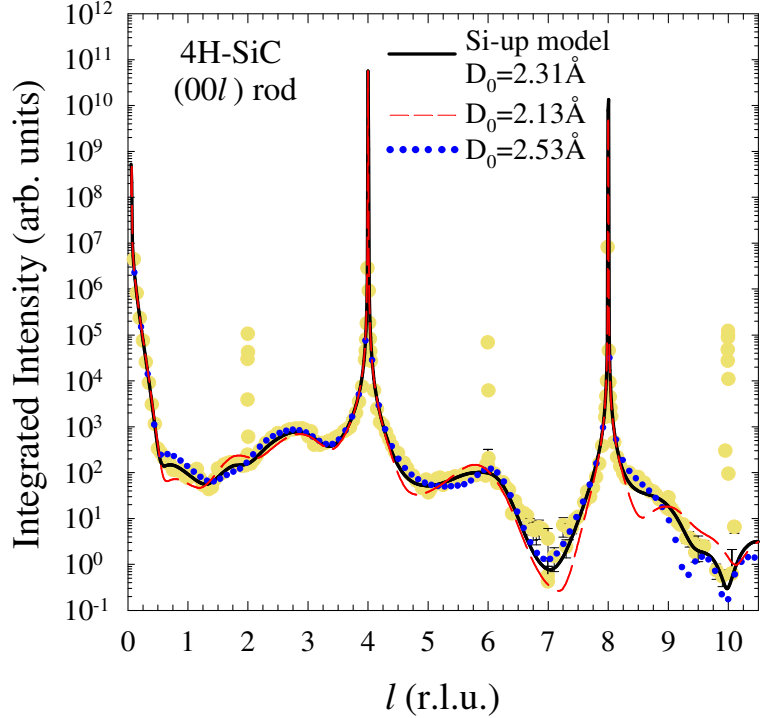}
\end{center}
\caption{(color online) Specular reflectivity versus $q_\perp$ (in r.l.u.) for a graphitized 4H-SiC(0001) Si-face surface.  Circles are the data. Solid black line is the best fit to the Si-up model in Fig.~\ref{F:Si-facemodel}(b) with $D_0\!=\!2.31$\AA (the C-adatom and Si-down model fits would be indistinguishable). Dashed red line is the same fit but with a smaller $D_0\!=\!2.13$\AA. Dotted blue line is the same fit but with a larger $D_0\!=\!2.53$\AA.}
\label{F:reflec_Do}
\end{figure}

\begin{table*}
\caption{\label{tab:Si-face_values}Best-fit structural parameters for graphite covered 4H-SiC$(0001)$ Si-face. Data for both the ``Si-up'', ``Si-down'' and ``C-adatom'' models give nearly identical fits. Parameters are defined in Fig.~\ref{F:Si-model_sch}}
\begin{ruledtabular}
\begin{tabular}{cccccccccccc}
&$d_{ad}$ (\AA)&$\rho_{ad}$ (\AA)&$\Delta_{1b}$(\AA)&$\rho_{1b}$ (\AA)
&$\Delta_{1a}$ (\AA) &$\rho_{1a}$ &$\Delta_{2C}$(\AA)&$\rho_{2C}$&$\Delta_{2Si}$(\AA) & $\rho_{2Si}$&\\
\hline Si-up& 0.88 & 0.23 & -0.23& 1.44 & -0.47 & 0.74 & -0.29 & 0.86 & -0.16& 0.81 &\\
Atom Type&   & silicon& & carbon & & carbon& &  &&&\\
\hline Si-down& 0.91& 0.58& -0.28& 1.50 & -0.48 & 0.26 & -0.28 & 0.86 & -0.16& 0.82 &\\
Atom Type&   & carbon& & carbon & & silicon& &  &&&\\
\hline C-adatom&  0.89 & 0.61 & -0.25& 1.38 & -0.48 & 0.71 & -0.28 & 0.85 & -0.15 & 0.81 &\\
Atom Type&   & carbon& & carbon & & carbon& &  &&&\\
\hline uncertainty&0.04& 0.08 & 0.04 & 0.08 &  0.04 & 0.1  & 0.05 & 0.1 & 0.05 & 0.08 &\\
\end{tabular}
\end{ruledtabular}
\end{table*}

The best reflectivity fit to the data is nearly identical for all three models and is shown in Fig.~\ref{F:reflec_Do}. Table~\ref{tab:Si-face_values} gives the fitting parameters for all models (uncertainty limits include variations from sample to sample). In the C-adatom model a carbon rich layer composed of three carbon layers is sandwiched between the graphene and a distorted SiC bilayer. The total density of these three interface layers is $\rho\!=\!0.61\!+\!1.38\!+\!0.71\!=\!2.70\!\pm 0.15$. This density is lower than the density of a graphene sheet ($\rho_\text{G}\!=\!3.13$) but is $30\%$ higher than a bulk SiC bilayer. The two Si models are similar to the C-adatom model in that they contain a carbon rich layer, although it is composed of two rather than three carbon layers sandwiched between the graphene. The total carbon density of the interface layers in the Si-up and Si-down models are $\rho\!=\!1.44+0.74\!=\!2.18\!\pm 0.15$, and $0.58+1.50\!=\!2.08\!\pm 0.15$, respectively. These values are similar to the total bulk bilayer density ($\rho\!=\!2.0$) needed to form a $\text{sp}^3$ bonded carbon layer. The two Si models are distinguished by a low density of Si atoms either atop or below the carbon rich interface layer.

There are two similarities between all three models. First, the high carbon densities in all three models suggest a complicated carbon bonding geometry that is neither like bulk SiC or like graphene. This has also been noted by both Emtsev et al.\cite{Emtsev_MSF_07,Emtsev_PRB_08} and Johansson et al.\cite{Johansson_PRB_96} who studied the \sixrt reconstruction that forms before graphitization and is known to persist after a true graphene layer has formed. The ARPES data of Emtsev et al.\cite{Emtsev_MSF_07,Emtsev_PRB_08} shows that the interface layer has $\sigma$ bands (although shifted to higher energy) but no $\pi$ bands. This suggests that the carbon concentration is high enough to at least locally support a $\text{sp}^2$ bonding geometry. In addition, the XPS studies of both Johansson et al.\cite{Johansson_PRB_96} and Emtsev et al.\cite{Emtsev_MSF_07} find two surface related C 1s core level shifts. As a result both studies conclude that the \sixrt surface contains a large amount of non-graphitic carbon in inequivalent surface sites in spite of the $\sigma$ bands.

The second similarity between these models is that, unlike the SiC$(000\bar{1})$ C-face,\cite{Hass_PRB_07} the Si-face interface reconstruction extends deeper into the bulk. The bilayer between the interface layer and the bulk is substantially altered from a bulk bilayer in both density and bonding. This deep reconstruction on the (0001) surface is consistent with the prediction of Johansson et al.\cite{Johansson_PRB_96} based on relative intensity ratios of surface to bulk XPS peaks. Regardless of the model, the bond lengths between this bilayer and both the bulk and the interface layer are contracted by $\sim\! 17\%$ from the bulk Si--C bond length making them similar to the bonds in diamond ($1.54\text{\AA}$).\cite{Burdett}.

\section{Discussion}\label{S:Disc}
The main difference between the ``C-adatom'' and either the ``Si-up'' or ``Si-down'' models is the low density Si layer in the interface region. While x-ray data alone cannot discriminate between an all carbon interface and a carbon rich interface with silicon, spectroscopic measurements strongly favor the two models with silicon in the interface. XPS and photoemission spectroscopy (PES) experiments conclude that, after graphene has formed, a significant fraction of Si remains at the interface.\cite{Johansson_PRB_96,Chen_SS_05,Ong_PRB_06} A complete XPS study by Johansson et al.\cite{Johansson_PRB_96} find that in addition to two surface-related C 1s core level shifts, there are also two surface-related Si 2p core level shifts.\cite{Note-on-Arpes} This is consistent with a Si adatom layer in the interface and a modified Si-C bond between the interface and the bulk-like bilayer below. We note that the bonding configuration of the Si-up model is very similar to a model proposed from STM images of the \sixrt interface structure below a layer of graphene.\cite{Rutter_PRB_07} The model of Rutter et al.\cite{Rutter_PRB_07} suggests an adatom density of $\rho_\text{ad}\!=\!0.22$; within error bars of the x-ray value in Table~\ref{tab:Si-face_values}. At the moment, however, there is no experimental data that can exclude either of the two Si models.

\begin{table*}
\caption{\label{tab:Si-C_compare}Structural parameters for graphene grown in UHV on 4H-SiC$(0001)$ Si-face (determined from this work) and those from furnace grown 4H-SiC$(000\bar{1})$ C-face graphene measured in Ref.~[\onlinecite{Hass_PRB_07}]. Parameters are defined in Fig.~\ref{F:Si-model_sch}}
\begin{ruledtabular}
\begin{tabular}{ccccccc}
 &$D_0$ (\AA)&$D_1$ (\AA)&$D_G$(\AA)&$\sigma_{G}$ (\AA)&$\gamma_\text{SiC}$&\\
\hline Si-Face& $2.32\pm 0.08$ & $3.50\pm 0.05$&$3.35\pm 0.01$ & $0.16 (-.05/+.02)$&$0.7\pm 0.1$&\\
C-face& $1.62\pm 0.08$& $3.41\pm 0.04$&$3.368\pm 0.005$ & $<0.05$ &$0.03\pm 0.01$&\\
\end{tabular}
\end{ruledtabular}
\end{table*}

It is worth pointing out a number of important structural differences between Si-face grown graphene and C-face grown graphene. Table~\ref{tab:Si-C_compare} shows a comparison of structural parameters determined from x-ray reflectivity data for graphene grown on the SiC(0001) and $(000\bar{1})$ surfaces. The distance between the first graphene layer and the interface for Si-face graphene film is $D_0\!=\!2.3\pm 0.08\text{\AA}$. Figure \ref{F:reflec_Do} shows the sensitivity of the fit to either increasing or decreasing the value of $D_0$ by $9\%$. The measured value of $D_0$ is large compared to the bilayer distance in bulk SiC ($1.89\text{\AA}$) and at the same time less than the graphite interplanar spacing of $3.354\text{\AA}$.\cite{Baskin_PR_55} Note that the best fit value of $D_0$ is similar to the value of 2.5\AA~ measured by STM.\cite{Rutter_PRB_07} The Si-face value of $D_0$ is larger than the value for furnace grown graphene on the SiC$(000\bar{1})$ C-face, implying that graphene is more strongly bound to the C-face interface. This is consistent with {\it ab initio} electronic calculations,\cite{Varchon_PRL_07,Mattausch_PRL_07} and the conclusions of previous inverse photoemission, PES and XPS studies\cite{Forbeaux_SS_99,Forbeaux00,Johansson_SS_98} (recent ARPES experiments have a contradictory interpretation to these studies, suggesting that the C-face graphene grown in UHV is in fact less tightly bound to the interface compared to the Si-face \cite{Emtsev_PRB_08}).  It is also worth pointing out that Table~\ref{tab:Si-C_compare} compares Si-face graphene to C-face graphene grown in a furnace; UHV C-face graphene used in previous studies\cite{Forbeaux_SS_99,Forbeaux00,Emtsev_PRB_08} grows nearly 200\degC~ lower in UHV compared to graphene grown in a furnace.\cite{Hass_JPhyCM_08} This temperature difference may influence the interface structure or order and should be considered when comparing results from different groups.

In addition to $D_0$, Table~\ref{tab:Si-C_compare} also shows that the spacing, $D_1$, between between graphene layer-1 and layer-2 is larger than the spacing between all other graphene layers (for both Si-face and C-face graphene).  Note that the error bar on $D_G$ is significantly larger than those reported for C-face graphene films.  This is because furnace grown C-face graphene films are much thicker than UHV grown Si-face films. Thinner films broaden the graphene Bragg peaks at $\ell\!=\!3$, 6, and 9, making the peak position and thus the layer spacing more difficult to measure. Table \ref{tab:Si-C_compare} also shows that the roughness of the SiC surface is more than an order of magnitude larger for Si-face graphene than for C-face graphene (the surface roughness is proportional to $\gamma_\text{SiC}$).  This is consistent with earlier x-ray results~\cite{Hass_APL_06} measuring growth-induced substrate roughness and with the substrate roughness observed in many STM and AFM images.\cite{de Heer_SSC_07,Seyller_SS_06,Riedl_PRB_07,Gu_APL_07}

The most important result of this work is that the interface layer for Si-face graphene is not a simple relaxed bulk termination of the SiC surface. This has a bearing on how to interpret electronic structure calculations of the graphene-SiC interface. To date, {\it ab initio} electronic structure calculations of the graphene/SiC(0001) system have started from a flat graphene layer placed above an idealized bulk terminated SiC surface that is allowed to relax into a slightly distorted bilayer.\cite{Varchon_PRL_07,Mattausch_PRL_07,Rutter_PRB_07} These calculations use an artificially contracted graphene sheet that is commensurate with a small SiC cell to allow for reasonable fast calculation times. Rutter et al.\cite{Rutter_PRB_07} have looked at a Si adatom model but, as with other calculations, used a simple relaxed bulk SiC bilayer below the adatoms. The result of all these calculations is that the first graphene layer above the relaxed bulk SiC bilayer acts as buffer that partially isolates the electronic properties of the next graphene layer from the substrate.\cite{Varchon_PRL_07,Mattausch_PRL_07} While these calculations are an important first step in predicting the existence of a buffer layer, their ability to predict the structure of the interface and thus its electronic properties is a concern given that the x-ray results show a much more substantial reconstruction that has few characteristics of a bulk SiC bilayer. Further experimental evidence for a buffer layer comes from ARPES, where a carbon-rich layer with substantial $sp^2$-bonding is found without any indication of $\pi$ bands characteristic of graphene.\cite{Johansson_PRB_96,Emtsev_MSF_07}  ARPES measurements also clearly show $\pi$-bonded graphene layers above this carbon-rich layer, although there are different interpretations of spectral structure very close to the K-point of the graphene Brillouin zone for the first of these structural graphene layers.\cite{Bostwick_NatPhys_07,Zhou_NatMat_07} We suggest that in the \emph{ab initio} results, the first graphene layer in the calculation mimics, to some extent, the properties of the interface layer-0. In fact recent {\it ab initio} calculations, using a full \sixrt cell, find that the first graphene layer is significantly distorted from a flat graphene sheet and does not show the dispersion characteristic of an isolated graphene sheet.\cite{Varchon_axCM_Ripples,Kim_PRL_08} In fact the calculated modulation amplitude is 1.23\AA, a value not far from the $\sim\!1.8$\AA~ interface layer width measured for all the structural models in Fig.~\ref{F:Si-facemodel}. These results suggest that the assumption of a distorted thick carbon-rich layer acting like a buffer layer may be correct.  A more realistic interface calculation will be necessary to test this assertion.

\begin{figure}[htbp]
\begin{center}
\includegraphics[width=10.0cm,clip]{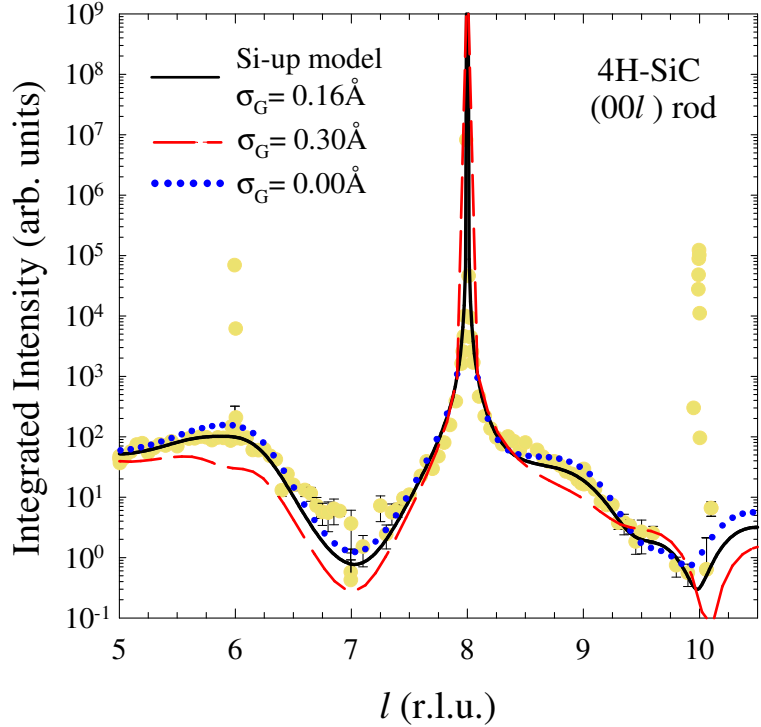}
\end{center}
\caption{(color online) Specular reflectivity versus $q_\perp$ (in r.l.u.) for a graphitized 4H-SiC(0001) Si-face surface.  Circles are the data. Solid black line is the best fit to the Si-up model with $\sigma_\text{G}\!=\!0.16$\AA. Dashed red line is the same fit but with $\sigma_\text{G}\!=\!0.30$\AA. Dotted blue line is the same fit but with a larger $\sigma_\text{G}\!=\!0.0$\AA. } \label{F:reflec_sigmao}
\end{figure}

In addition to the interface structure, the reflectivity data provides additional evidence supporting conclusions based on STM and LEED that the \sixrt reconstruction observed after graphitization is a true reconstruction of the graphene film. Since the discovery of the \sixrt LEED pattern, it has been suggested that it is a moir\'{e} pattern due to the near commensuration of graphene with SiC.\cite{vanBommel75,Chang_SS_91,Li_SS_96} Early STM experiments supported this claim because they imaged a $6\!\times\!6$ reconstruction\cite{Chang_SS_91,Owman_SS_95,Li_SS_96,Tsukamoto_ASS_97,Chen_SS_05} instead of the \sixrt pattern observed in LEED. More recent STM experiments on the other hand, have directly imaged the \sixrt structure and shown that the graphene has a vertical modulation with this lateral periodicity.\cite{Rutter_PRB_07,Riedl_PRB_07} Recent {\it ab initio} calculations, using a full \sixrt cell, find that at least the first graphene layer above the SiC has a substantial modulation amplitude.\cite{Varchon_axCM_Ripples} The x-ray reflectivity data supports these recent experiments and confirms that there is a vertical modulation of graphene grown on the Si-face. This can be seen by comparing the C-face and Si-face graphene layer roughness or corrugation, $\sigma_G$, from Eq.~(\ref{E:Fpb}) [see Table~\ref{tab:Si-C_compare}]. $\sigma_G$ is much larger on Si-face grown graphene than on C-face furnace grown graphene. $\sigma_G$ is determined almost solely by the intensity decay of the graphite Bragg points as a function of $\ell$. Because of the exponential form in Eq.~(\ref{E:Fpb}), a finite $\sigma_G$ manifests itself as a decay in the graphite Bragg peak intensities at $\ell\!=$6 and 9. This is demonstrated in Fig.~\ref{F:reflec_sigmao} where we compare fits from a graphene film with no roughness ($\sigma_G\!=\!0$\AA) to a fit with a large roughness ($\sigma_G\!=\!0.3\text{\AA}$) and to the best fit value for the Si-up mode in these experiments $\sigma_G\!=\!0.16$\AA. Note in Table \ref{tab:Si-C_compare} that $\sigma_G$ is essentially zero for C-face graphene films that show no LEED reconstruction patterns.\cite{vanBommel75,Forbeaux_SS_99,Forbeaux00,Hass_APL_06} We can interpret $\sigma_G$ as originating from an actual modulation of the graphene film, but the value of 0.16\AA~is considerably smaller than the value of 0.6\AA~measured by STM for the first graphene sheet above the interface layer.\cite{Rutter_PRB_07} This difference arises in part because STM is measuring a modulation in the electron density of states instead of an actual structural modulation but more importantly, because the x-ray value is an average over all the graphene layers in the film. Riedl et al.\cite{Riedl_PRB_07} have shown that the vertical modulation amplitude decays by approximately a factor of two from the first to the second graphene layer. Therefore, thicker graphene films weight the average $\sigma_G$ to lower values.

\begin{figure}
\begin{center}
\includegraphics[width=10.0cm,clip]{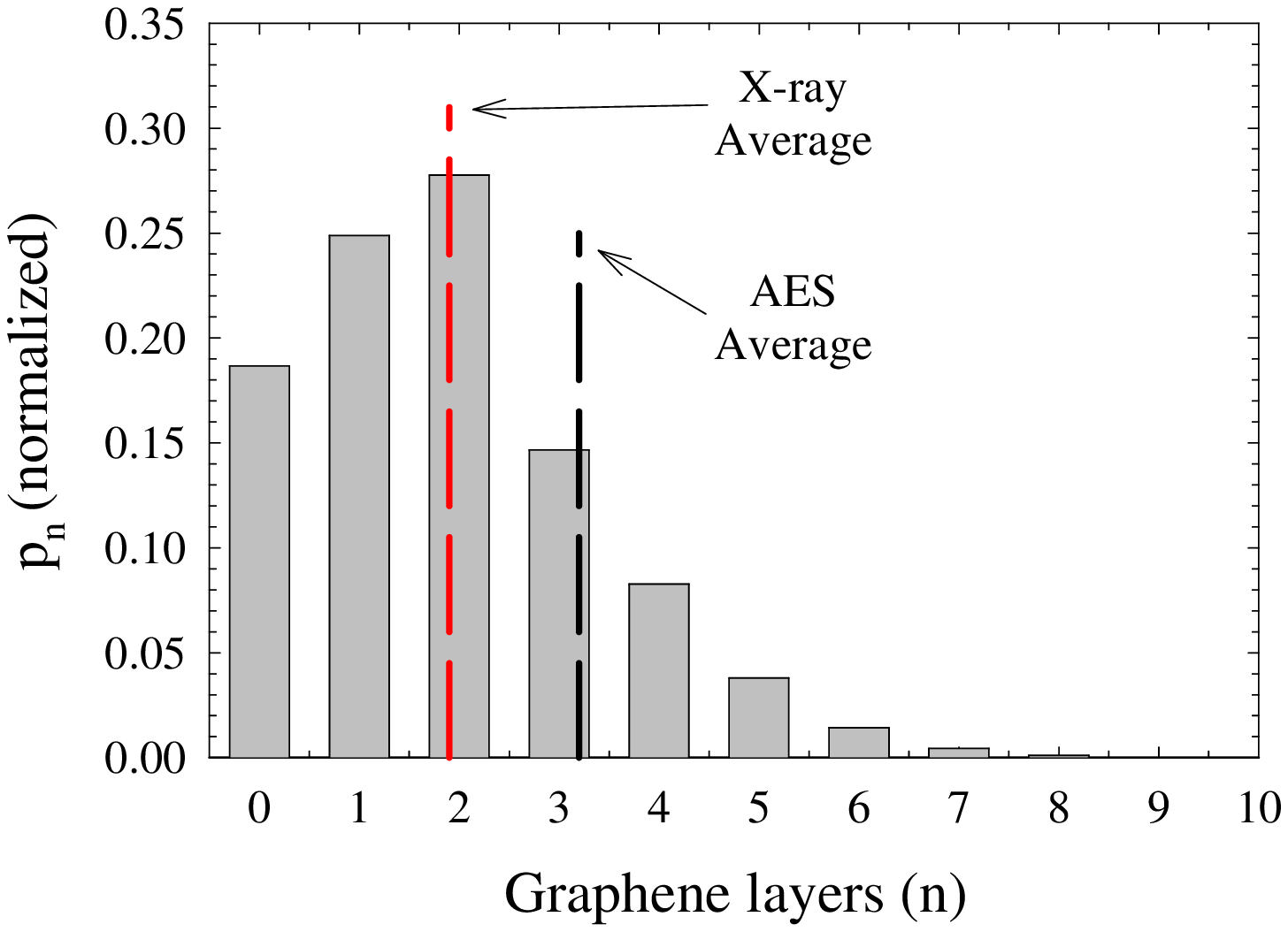}
\caption{The normalized probability, $p_n$, of a $n$-graphene layer stack from a UHV grown Si-face film as determined by x-ray reflectivity. The x-ray average is $1.9\!\pm\!1.5$ graphene layers while the AES estimate of the average is approximately one layer thicker (3.2 layers).}
\end{center}
\label{F:Si-face_Histo}
\end{figure}

We can estimate the first graphene layer modulation, $\sigma_G^{(0)}$, from the measured mean modulation $\sigma_G$ if we assume that the modulation decays in subsequent layers as $\sigma_G^{(n)}\!=\!\sigma_G^{(0)}\exp[-\lambda Dn]$ where $\lambda D=\ln(2)$ (the factor of $\ln(2)$ assumes the amplitude decay measured by Riedl et al.\cite{Riedl_PRB_07} is correct).  To calculate $\sigma_G^{(0)}$ we only need to know the relative amount of graphene that is thicker than $N$ layers, $P_N$. $P_N$ is calculated from the areal coverage, $p_n$'s in Eq.~(\ref{E:Fpb}); $P_N\!=\!C\sum_{n=N}^{N_{max}}{p_n}$ (C is a normalization constant). Then $\sigma_G^{(0)}$ is given by;
\begin{equation}
\sigma_G^{(0)}=\sigma_G\sum_nP_n/\sum_nP_n\exp{[-\lambda Dn]}.
\label{E:sig_o}
\end{equation}
The measured distribution of graphene layer thickness, $p_n$, for a nominally 2-layer graphene film is shown in the layer height histogram in Fig.~\ref{F:Si-face_Histo}. The average number of graphene layers is $1.9\pm\!1.5$. The distribution is very wide, in part reflecting the spatial average over the large x-ray beam footprint (the footprint is bigger than the sample width of 3mm when $\ell\!<\!1.8$). In particular the high areal fraction not covered by graphene ($18\%$) can be associated with slow growth kinetics at cooler substrate regions near the edge of the sample caused by non-uniformity in the e-beam heating. These wide distributions are also seen in low energy electron microscope (LEEM) images.\cite{Hibino_PRB_08,Ohta_LEEM} Using this measured distribution, we estimate $\sigma_G^{(0)}$ to range between 0.5-0.8\AA reflecting the uncertainty in the measured value of $\sigma_G$. This result is in very good agreement with the STM value.\cite{Rutter_PRB_07}

For comparison, we can estimate the vertical modulation of the first graphene layer using a different method. This is done by using a modulated first graphene layer on-top of the Si-up adatom model to calculate $F_1(\ell)$ in Eq.~(\ref{E:Fpb}).  To do this we use the graphene structural coordinates calculated by Varchon et al.\cite{Varchon_axCM_Ripples} The modulation of higher graphene layers are included using the rms roughness in the standard model. While the calculated relative vertical carbon positions are maintained in the first graphene layer, the absolute positions are scaled by a multiplicative constant so that the peak-to-peak amplitude can be varied. The best fit structure to the reflectivity data gives the first graphene layer amplitude to be 0.82\AA peak-to-peak, which is slightly larger than than the range of $\sigma_G^{(0)}$ estimated above.

It is worth comparing the graphene thickness measured by SXRD and an estimate from the simpler AES method.\cite{Muehlhoff_JAP_86,Tsukamoto_97,T_Li_thesis} The ratio of the Si(LVV)/C(KLL) peak area can be used, along with proper electron mean free paths and excitation cross sections, to estimate of the number of graphene layers provided a suitable model for the interface is used. Most groups use an interface model consisting of a bulk terminated substrate with graphene above.\cite{Muehlhoff_JAP_86,Tsukamoto_97} (Ref.~\onlinecite{T_Li_thesis} considers a $\sqrt{3}\times\sqrt{3}$ Si adlayer). While bulk termination is clearly inconsistent with both models in Fig.~\ref{F:Si-facemodel}, it can still be used with the proviso that it will overestimate the film thickness by $\sim\!1$ layer because the measured C(KLL) intensity includes a contribution from the dense non-graphitic interface carbon layer in Fig.~\ref{F:Si-facemodel}. AES measurements on the same sample as the data for Fig.~\ref{F:Si-face_Histo} estimate the average graphene coverage to be 3.2 layers compared to 1.9 by x-rays, consistent with the lack of a realistic interface layer in the AES calculation. Because the electron penetration depth of the Si(LVV) electron is short, AES estimates become more uncertain for graphene layers exceeding four layers. This makes the AES method more applicable to Si-face films than to thick furnace grown C-face graphene films.

\section{Conclusion}
We have used surface x-ray reflectivity measurements to determine the graphene/SiC interface structure for graphene grown on the 4H-SiC(0001) Si terminated surface (Si-face). We find that the interface is not composed of a simple graphene-like layer above a relaxed SiC bilayer that has been recently proposed.\cite{Emtsev_PRB_08} Instead, the interface reconstruction is more complicated and extends deeper into the bulk.  Based on these x-ray experiments and previous XPS studies~\cite{Johansson_PRB_96} the best model for the interface is composed of a substantially relaxed SiC bilayer, above which, a dense carbon layer containing a partial layer of Si atoms separates it from the graphene film. The carbon density in this intermediate layer is approximately 2.1 times larger than in a SiC bilayer. This model is consistent with previous STM~\cite{Owman_SS_95,Rutter_PRB_07,Riedl_PRB_07} and XPS~\cite{Johansson_PRB_96} results as well as the lack of interface $\pi$ bands in ARPES experiments.~\cite{Emtsev_MSF_07,Emtsev_PRB_08} The bond distance between the Si adatom layer and the first graphene layer (Si-up model) is $2.32\pm\!0.08$\AA. While this distance is short compared to the interplanar graphene spacing, it is still larger than the corresponding distance measured on furnace grown SiC$(000\bar{1})$ C-face graphene ($1.62$\AA),\cite{Hass_PRB_07} indicating that the graphene on the Si-face is less tightly bound to the substrate than furnace grown C-face graphene. This difference is consistent with {\it ab-initio} calculations.\cite{Varchon_PRL_07} We propose that the dense carbon layer with Si adatoms plays the role of the buffer layer predicted by Varchon et al.\cite{Varchon_PRL_07}  It is this layer that partly isolates subsequent graphene layers from interactions with the substrate.

The x-ray results also indicate that the graphene is not flat but has a corrugation amplitude consistent with that measured by STM.\cite{Rutter_PRB_07,Riedl_PRB_07} This result confirms that the \sixrt periodicity imaged by STM is dominated by a real structural corrugation of the graphene and not by changes in the local density of states.

\begin{acknowledgments}

We wish to thank I. Gadjev and K. Charulatha  for helping to prepare the reflectivity data and fits. We also would like to thank C. Berger for helpful discussions and F. Varchon and L. Magaud for providing the atomic coordinates from their graphene/SiC interface calculations. This research was supported by the National Science Foundation under Grant Nos. 0404084 and 0521041, by Intel Research and by the W.M. Keck Foundation. The Advanced Photon Source is supported by the DOE Office of Basic Energy Sciences, contract W-31-109-Eng-38.  The $\mu$-CAT beam line is supported through Ames Lab, operated for the US DOE by Iowa State University under Contract No.W-7405-Eng-82.
\end{acknowledgments}


\end{document}